\documentstyle[epsfig,prb,aps]{revtex} 
\tighten 
\begin{document} 
\sloppy 
\twocolumn[
\hsize\textwidth\columnwidth\hsize
\csname @twocolumnfalse\endcsname

\title{Multiple Andreev Reflection and Giant Excess Noise in
Diffusive Superconductor/Normal-Metal/Superconductor Junctions} 

\author{T.~Hoss, C.~Strunk,\cite{cs} T.~Nussbaumer, R.~Huber,\cite{rh} U.~Staufer,\cite{us} and C.~Sch\"onenberger}
\address{Institut f\"ur Physik, Universit\"at Basel,
Klingelbergstr.~82, CH-4056 Basel, Switzerland}

\date{February, 10$^{th}$, 2000}

\maketitle

\begin{abstract}

We have studied superconductor/normal metal/superconductor  (SNS)
junctions consisting of short Au or Cu wires between Nb or Al banks. The Nb based junctions display inherent electron heating effects induced by the high thermal resistance of the NS boundaries.
The Al based junctions show in addition subharmonic gap structures in the
differential conductance $dI/dV$ and a pronounced peak in the
excess noise at very low voltages $V$. We suggest that the noise peak is caused by fluctuations of the supercurrent at the onset of Josephson coupling between the superconducting banks. At intermediate temperatures where the supercurrent is suppressed a noise contribution $\propto 1/V$ remains, which may be interpreted as shot noise originating from large multiple charges.

\end{abstract} 

]

\section{Introduction}
Superconductor/normal-metal (SN) interfaces of high
transparency exhibit remarkably different properties for electric charge
and energy transfer, respectively. Quasiparticles with energy $\epsilon$ below the energy gap $\Delta$
of the superconductor cannot enter the superconductor. This implies a
high thermal resistance of the SN boundary since energy is
exclusively carried by the quasiparticles.\cite{andreev} In contrast,  
charge can be transmitted at $\epsilon < \Delta$ via
the Andreev reflection process: An electron coming from the normal 
side is reflected as a hole and a
Cooper pair is transferred to the superconductor.\cite{btk}  This
should have important consequences for the energy distribution of
quasiparticles in a short normal bridge connected to two superconducting
reservoirs. Here we assume that the length $L$ of the bridge is
larger than the thermal diffusion length $L_T=\sqrt{\hbar D/2\pi k_BT}$
which governs the penetration of Cooper pairs into the normal wire.
For $L>L_T$ the supercurrent through the structure is 
exponentially weak. 

In the case of normal reservoirs the distribution function $f(\epsilon)$ in the wire 
either assumes a two step shape if the inelastic scattering length
$L_{in}(\epsilon) \gg L$, or smears out into a Fermi-Dirac function with a
spatially varying electron temperature
if $L_{in}(\epsilon) \ll L$.\cite{nagaev92,poth97,stei96} The
broadening of $f(\epsilon)$ has been detected by local tunneling spectroscopy,\cite{poth97} or by measuring the power spectral density 
$S_V(V)$ of the current noise in the junction.\cite{stei96,schoel97,henny99}

In the case of superconducting reservoirs the broadening of $f(\epsilon)$
is expected to be much more dramatic when compared to normal reservoirs. 
In particular for small applied
voltages $eV\ll \Delta$ quasiparticles have to climb up to the energy gap
$\Delta$ via multiple Andreev reflections (MAR) at the two SN boundaries in order to remove the deposited energy into the
reservoirs. For samples shorter than the phase coherence length $L_\phi$ 
subharmonic gap structures have been observed in diffusive samples\cite{kutch97}  in the differential conductance $dI/dV$ at voltages close to
$V=2\Delta/en$ where $n$ counts the number of
reflections in the MAR cycle. Such structures have been found earlier in superconducting microbridges, tunnel junctions and ballistic S/N/S point contacts.\cite{greg73} In addition, there are theoretical\cite{aver96,cuev98,naveh99} and experimental\cite{diele97} indications that the coherent MAR cycle  transfers multiple charge quanta
of magnitude $2\Delta /V$, which should lead to an enhanced current
noise at low bias voltages. For diffusive systems so far only indications of charge doubling have been reported, which points towards single Andreev reflection events.\cite{jehl99}

In this work, we address the above questions by measuring $dI/dV$ 
and $S_V$ of high transparency Nb/Au/Nb, Al/Au/Al and Al/Cu/Al junctions 
prepared by means of a novel anorganic shadow mask.\cite{mne} 
Compared to previous studies\cite{cour95} we focus on junctions having a very small critical current. This allows to study the low voltage regime which is otherwise difficult to access because of heating effects. 

\section{Sample preparation}
The samples are prepared by angle evaporation through a suspended Si$_3$N$_4$ mask on a Si substrate with a SiO$_2$ spacer layer.\cite{mne} The anorganic mask avoids the previously observed deterioration of the
superconducting properties of the Nb by outgassing of the conventional organic resist
(e.g.~PMMA) during evaporation of the high melting point Nb.\cite{hara94} 
The Si$_3$N$_4$ top layer is patterned by conventional electron beam lithography. 
Wet etching of the 800 nm SiO$_2$ layer results in the desired undercut profile. 
The high mechanical strength of the Si$_3$N$_4$ allows large undercuts and
freely suspended brigdes of several micron length. The transition temperature  of  
narrow Nb wires displayed only a minor reduction (0.2 K) of the superconducting 
transition temperature $T_c$ when compared to codeposited Nb films.\cite{mne} 
Our SNS devices consist of thin ($\simeq$ 15 nm) normal wires (Au or Cu) of
0.4 - 2 \mbox{$\mu$m} length and 100 - \mbox{200 nm} width between thick (50 - \mbox{200 nm}) reservoirs made of Nb or Al. A scanning elec-\begin{figure}[t]
\centerline{\psfig{file=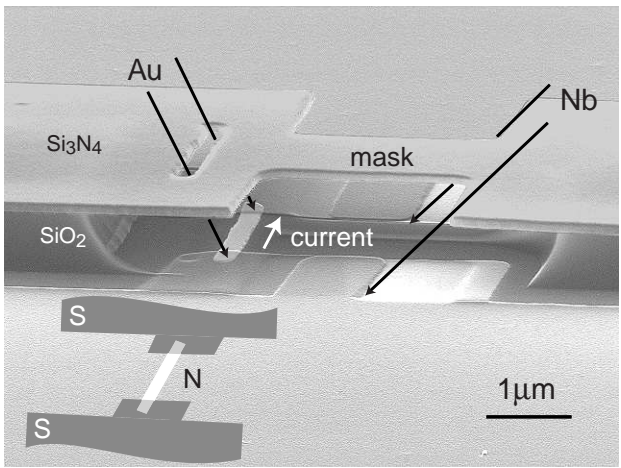,width=8.2cm}}
\bigskip\bigskip
\caption{\label{bridge} Scanning electron micrograph of a typical sample viewed under a large tilt angle. The normal wire between the two Nb reservoirs (top and bottom) is defined through the slit in the freely suspended nitride mask. Inset: schematic of the sample layout.} 
\end{figure} 
\noindent tron micrograph of a Nb/Au/Nb sample is shown in Fig.~\ref{bridge}.  

A single SNS junction and a series of 9 - 16 junctions together with \mbox{20 $\mu$m} long N and S wires were prepared simultaneously on the same chip. From weak localization measurements of the long wires we determined $L_{\phi}$ of the N metal. The spectral density $S_V$ of the voltage fluctuations across the sample is measured as a function of current bias in the frequency range between 100 and 400 kHz with a cross-correlation technique.\cite{henny99} With this technique we obtain a voltage sensitivity of \mbox{50 pV$^2/\sqrt{\mbox{Hz}}$} with commercial room temperature preamplifiers.
The measurements are performed in a $^3$He cryostat which is shielded from rf interference by $\pi$ filters at room temperature and by a thermocoax filtering stage at the \mbox{0.3 K} stage. The sample is put into an rf tight copper chamber at the sample temperature.

\section{Nb/Au/Nb Junctions}
\begin{figure}[p]
\centerline{\psfig{file=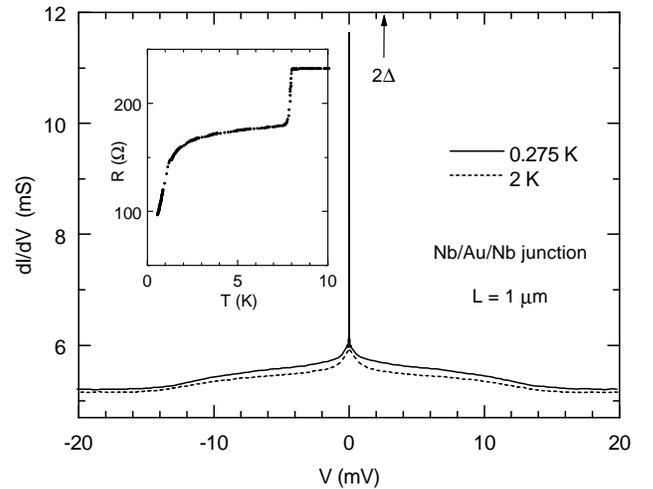,width=8.2cm}}
\bigskip\bigskip
\caption{\label{dvdinb} Differential conductance $dI/dV$ of a single Nb/Au/Nb junction as a function of voltage $V$ for two temperatures. The sharp peak at $V=0$ indicates the rapid destruction of electron-hole coherence by a finite bias voltage. The arrow indicates 
$V=2\Delta/\mbox{e} = \,$\,\mbox{2.6 mV}. Note the absence of  subharmonic gap structures. 
The Au wire is \mbox{1 $\mu$m} long, \mbox{130 nm} wide and \mbox{15 nm} thick. The thickness of the Nb reservoirs is 50 nm. Inset: Resistance vs.~temperature for the same sample. }
\end{figure}

In Fig.~\ref{dvdinb} we show the differential conductance $dI/dV$ vs.
applied voltage of a single Au wire of length \mbox{$L=1\,\mu$m} between Nb
reservoirs. The inset displays $R(T)$ for the same sample. The data are recorded using ac currents of typically 10 and \mbox{20 nA}. When
lowering the temperature the resistance first drops around \mbox{8.2 K} which indicates the superconducting transition of the reservoirs. 
Further reduction of the temperature leads to a continuous decrease of $R$
which becomes more drastic below 2 K. 
This proximity induced reduction of $R$ is accompagnied by a sharp peak in the differential conductance $dI/dV$ at zero bias voltage. The peak has a width of \mbox{50 $\mu$V} (which is close to $k_BT$ at \mbox{0.3 K}) and can be seen as the precursor  
\noindent of a supercurrent which emerges when $L_T$ becomes comparable to the  wire length $L$.
The peak has a height of only \mbox{10 \%} of the normal state resistance $R_N$ of the wire for \mbox{$L=2$ $\mu$m}, while we find supercurrents up to 50 $\mu$A for \mbox{$L=0.4$ $\mu$m} in samples with Nb banks.

\begin{figure}[p]
\centerline{\psfig{file=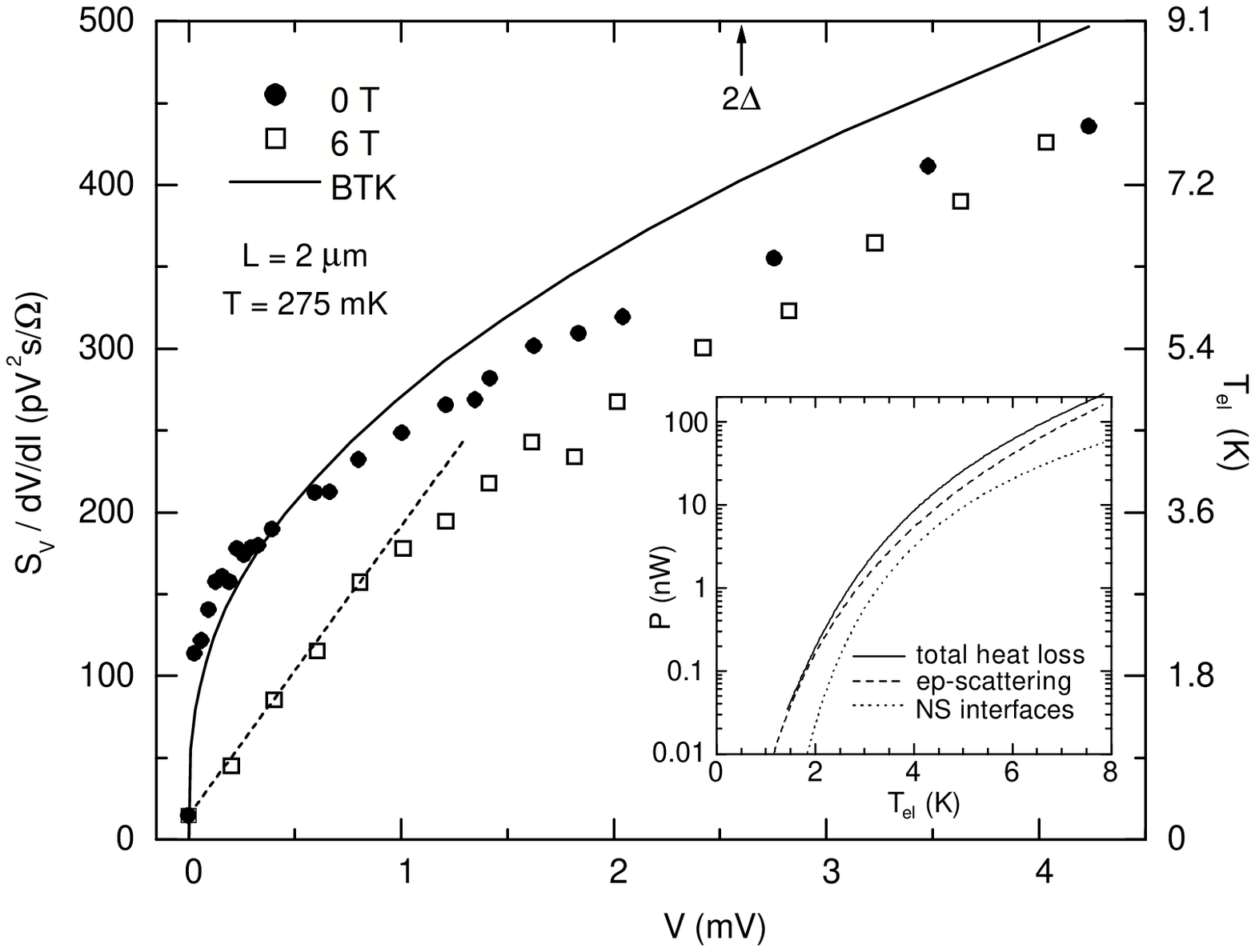,width=8.6cm}}
\bigskip\bigskip
\caption[]{\label{sinb} Scaled excess noise $S_V / dV/dI$ as a function of voltage for a series of 9 Nb/Au/Nb junctions of \mbox{2 $\mu$m} length, \mbox{200 nm} width and \mbox{15 nm} thickness for normal ($\Box$) and superconducting ($\bullet$) Nb reservoirs. The arrow indicates 
$V=2\Delta/\mbox{e} = \,$\mbox{2.6 mV}. The zero bias resistance per junction is \mbox{39 (72) $\Omega$} at \mbox{$B=0$ (6) T} and the diffusion constant of the wire is \mbox{$D\simeq$ 68 cm$^2/$s}. At \mbox{6 T} the Nb reservoirs contribute significantly to the junction resistance. The right-hand scale indicates the effective electron temperature. The dashed line indicates the shot noise of noninteracting electrons in case of normal reservoirs. The solid line gives an estimate of the electron heating effect according to Eqs.~(1) and (2). Inset: Electron-phonon (dashed line) and N/S interface (dotted line) contributions to the cooling power as a function of electron temperature $T_{el}$ in the wire. $T_{el}$ in the reservoirs is assumed to remain at \mbox{0.27 K}.} 
\end{figure}

In Fig.~\ref{sinb} we present the excess noise $S_V$ of a series of 9 Nb/Au/Nb junctions with 2 $\mu$m long Au wires.  As a reference measurement,
we first collected data in a perpendicular magnetic field of
\mbox{6 T} in which the Nb reservoirs are normal (open squares). For a direct comparison of the (effective) electron temperatures $T_{el}$ we have normalized $S_V$ with $dV/dI$ (see right-hand scale). For lower voltages \mbox{$V<$
1 mV} the measured noise falls on the \mbox{$1/3$} reduced shot noise (dashed line). At higher voltages, additional cooling via electron-phonon scattering results in a negative curvature of $S_V$.\cite{henny97} 

In the case of superconducting reservoirs (solid circles) we find a dramatic increase of $S_V$ in particular for the smallest voltages. The normalized excess noise
rises with nearly vertical slope at $V=0$ and merges at \mbox{$V\sim 2\Delta/e$} into the \mbox{6 T} curve. The latter is expected because for energies $\epsilon \gtrsim \Delta$ the probability of Andreev reflection rapidly vanishes. Note that $T_{el}$ is already \mbox{$\simeq$ 6 K} for \mbox{$V\,\simeq 2\Delta /e$}.
From weak localization measurements on the long Au wire we infer $L_{\phi} =$ \mbox {0.9 $\mu$m} at \mbox{1.3 K} (\mbox{0.55 $\mu$m} at \mbox{4.2 K}). Since $L_\phi$ is considerably shorter than the wire length of \mbox{2 $\mu$m} the MAR cycle is incoherent. This is confirmed by the absence of subharmonic gap features in $dI/dV$ (see Fig.~\ref{dvdinb}).

The electron temperature in the Au wire is controlled by the power dissipation in the wire, the energy loss via 

\noindent quasiparticle transmission through the S/N interfaces, and the electron phonon scattering in the Au wire.\cite{henny99}  At low $T_{el}$ the electronic heat diffusion within the Au wire is much faster than the energy loss across the interfaces so that
 we may assume local thermal equilibrium with a nearly constant temperature profile along the wire. The heat transfer through the interfaces can be reasonably well described in terms of a simple BTK-like expression\cite{btk} for the heat current $P_{NS}(T_{el})$ through the N/S boundaries:
\begin{equation}
P_{NS}(T_{el})=\frac{2}{R_me^2}\,\int\limits^\infty_{-\infty} \epsilon\,\,
(f_N-f_S)\,(1-A-B)\,d\epsilon\,\,\,.
\end{equation}
Here, $R_m$ is the normal state resistance of the N/S boundary, $f_N(T_{el})$ and $f_S(T_{Bath})$ are the Fermi functions in the wire and the reservoirs, while $A(\epsilon,Z)$ and $B(\epsilon,Z)$ are the coefficients of Andreev- and normal reflection and $Z$ is the interface parameter. We estimate $R_m \simeq$ \mbox{5 $\Omega$}. Within our simplified model, the cooling via electron phonon contribution scattering is given by 
\begin{equation}
P_{ep}(T_{el}) = \left(\frac{k_B}{e}\right)^2\frac{L^2\Gamma}{R_N}\,(T^5_{el}-T^5_{Bath})\,\,,
\end{equation} where $L$ is the length of the normal wire and $\Gamma\simeq 5\cdot10^8$ K$^{-3}$m$^{-2}$ for Au\cite{henny99}. 
The parameter $\Gamma$ is related to the electron-phonon scattering rate: 
$\tau^{-1}_{ep} = \zeta(3)/2\zeta(5)\,D\Gamma\,T^3_{el}$, where $\zeta(n)$ is the Riemann Zeta-function.\cite{wellstood} 
The calculated cooling power according to Eqs.~(1) and (2) is plotted as a function of $T_{el}$ in the wire (dotted and dashed line) in the inset of Fig.~\ref{sinb}. For simplicity  we assume $Z=0$. The solid line is the sum of both contributions and corresponds to the solid line in the main figure \ref{sinb}. Finite values of $Z$ lead to a shift of the solid lines to lower cooling power and to higher electon temperatures, respectively. 
At intermediate temperatures both contributions are of comparable magnitude, while the electron-phonon term wins at low temperature because of the exponential cut-off of the N/S interface term and at high temperature because of the strong $T_{el}^5$- increase of the electron phonon term.
In our geometry where the area of the N/S interface is tiny (200 $\times$ 200 nm$^2$), the N/S interface term is much smaller than in the related experiment on Nb/Al/Nb junctions by Jehl {\it et al.}, \cite{jehl99} who used subtractive structuring of a Nb/Al-bilayer. This may be the reason, why heating effects appear to be negligible in the latter experiment.

\section{Al/Cu/Al - Junctions}
\begin{figure}[p]
\centerline{\psfig{file=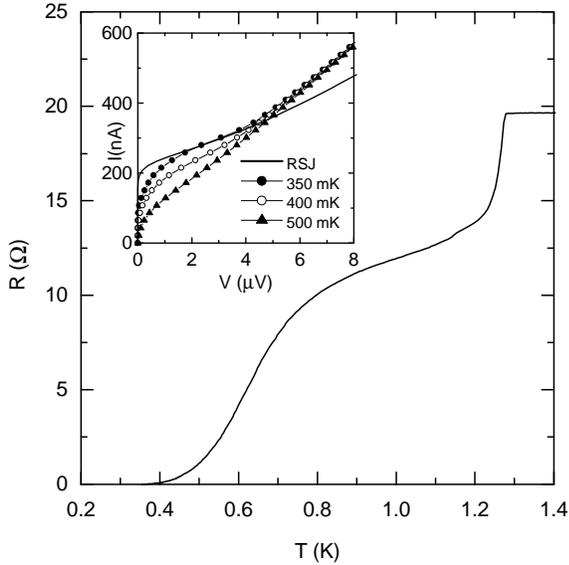,width=8cm}}
\bigskip\bigskip
\caption[]{\label{rtal} Resistance vs.~temperature of a series of 16 Al/Cu/Al junctions.  The Cu wires are \mbox{0.9 $\mu$m} long, \mbox{160 nm} wide and \mbox{18 nm} thick and have a diffusion constant \mbox{$D\simeq$ 72 cm$^2/$s}. The thickness of the Al reservoirs is \mbox{150 nm}. Inset: Current-voltage characteristics in the low voltage region. The symbols are the experimental data and thick the solid line is a fit according to the RSJ model using $I_c\simeq 270 $nA.\cite{amb69} } 
\end{figure}

It is now very interesting to look at samples, in which $L_{\phi}(\epsilon)$ remains larger than the wire length.  To avoid inelastic scattering, it is necessary to keep $T_{el}$ below $\simeq$ \mbox{1 K} in the voltage range $V \leq 2\Delta$. In a second
set of experiments we replaced Nb by Al, having a much
smaller gap $\Delta_{Al}$. As a consequence, the energies acquired in
the MAR cycle are much lower and we expect to enter the regime of
coherent Andreev reflection. For the normal wire we used  both Au and Cu, where for Cu we measured a longer phase coherence length of
\mbox{1.35 $\mu$m} at \mbox{1.3 K} 
than for Au. Figure \ref{rtal} shows the resistance vs. temperature of a series of 16 $\times$ \mbox{1 $\mu$m} long Al/Cu/Al junctions. When lowering $T$ the resistance sharply drops at the transition of the reservoirs \mbox{$\simeq$ 1.25 K} and then continuously vanishes as the proximity effect drives the Cu wire into a superconducting state. 
This sample shows zero resistance at the lowest $T$ since the Thouless energy $E_c=\hbar D/L^2 \simeq\,$5$\mu$eV and the normal state conductance are larger compared to the Nb/Au/Nb junctions. According to the theory by Wilhelm {\it et al.}\cite{cour95} the critical current $I_c(T)$ reads in the limit $k_BT \gg E_c$ :
\begin{equation}\label{Ic}
I_c(T) = \frac{3.0\,\mbox{mV/K}}{R_N\,\sqrt{T_0}}\,T^{3/2}\,\exp\left ( -\sqrt{T/T_0}\right )\,\,,
\end{equation}
where $T_0=E_c/2\pi k_B$.

The inset of Fig.~\ref{rtal} displays the measured current-voltage ($IV$) characteristics of the same sample. The turning point of the $IV$ curves indicates $I_c(350\,\mbox{mK}) \approx 300\,$nA. With our sample parameters we estimate from 
\noindent Eq.~\ref{Ic} a critical current $I_c(350\,\mbox{mK}) \simeq 510\,$nA.
This estimate is reasonably close to the measured values. On the other hand, we observe a substantial broadening of the transition such that the zero voltage state is reached only for currents $\lesssim$ \mbox{80 nA}.

At finite temperatures a certain intrinsic broadening of the $IV$ curves is expected by virtue of thermally activated phase slips which is usually described within the RSJ model.\cite{amb69} Our SNS junctions are self shunted with $R_N$ as the shunt resistance. We observe a broadening which is much stronger than expected from the RSJ model. This is illustrated by an RSJ fit using $I_c = 270\,$nA and $T=350\,$mK which is represented by the thick solid line in the inset of Fig.~\ref{rtal}.  In principle such an enhanced broadening can be caused by external electromagnetic interference. At high frequencies this source of broadening is suppressed by our rf filtering at room temperature and the sample stage. At lower frequencies we have checked, that the highest spikes in the frequency spectrum correspond to current noise below \mbox{1 nA/$\surd$Hz}, which is much lower than the critical current at \mbox{350 mK}.
We are therefore confident that there is an intrinsic origin of the broadening of the $IV$ curves.   At voltages \mbox{$V\gtrsim5\,\mu$V} the measured currents become larger than the fit. This is caused by the excess current induced by the Andreev reflection (see the discussion below).

Being made for tunnel junctions, a failure of the RSJ for long SNS junctions is not too surprising since it takes into account only the phase degree of freedom of the pair 
\begin{figure}[p]
\centerline{\psfig{file=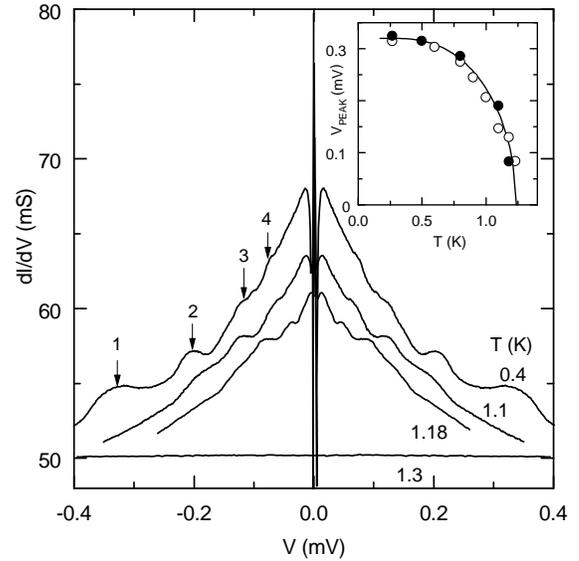,width=7.4cm}}
\bigskip\bigskip
\caption{\label{didval} Differential conductance $dI/dV$ vs.~voltage $V$ of the same sample as in Fig.~\ref{rtal} for several temperatures. The arrows indicate subharmonic gap structures corresponding approximately to integer fractions of $2\Delta$. Inset: Position of the $2\Delta$ conductance peak vs.~temperature for two samples with different normal state conductance (\mbox{$\bullet$ : 50 mS}, \mbox{$\circ$ : 29.3 mS}) . The solid line is a BCS fit for  2$\Delta$ =\mbox{325 $\mu$eV} and $T_c$ = \mbox{1.23 K}.} 
\end{figure}

\noindent amplitude $F(x)=\langle\psi_\downarrow\psi_\uparrow\rangle$, while it neglects spatial variations of the absolute value $|F(x)|$. In SNS junctions there is a minimum of $|F(x)|$ at the center of the N-wire. The minimum value of $|F(x)|$ at this 'weak spot' strongly depends on the ratio $L/L_T$ which is reflected in the temperature dependence of $I_c$ at temperatures $k_BT \ll \Delta$ described by Eq.~\ref{Ic}.
We believe that the enhanced rounding of the IV curves is related to the presence of the weak spot in the N-wire, which greatly facilitates phase slips in long SNS junctions. Correspondingly, also the shape of R(T) cannot be fitted with the RSJ formulas, since the temperature dependence of $I_c$ is superimposed on that of the thermal activation process. In particular, $R(T)$ does not follow a simple Arrhenius law. Broadened transitions induced by phase slip processes also occur in microbridges and long filaments made from homogeneous superconductors. The latter examples differ from SNS junctions in that the main temperature dependence comes from $\Delta(T)$, which is not important at temperatures $T\ll T_c$. 

The $dI/dV$ curves of the same sample but in a larger voltage range are presented for various temperatures in Fig.~\ref{didval}. Besides the supercurrent at $V=0$ we find a considerable conductance enhancement for $V<2\Delta$. In addition, conductance peaks close to $V = 2\Delta_{Al}/ne$ are present, which we attribute to coherent MAR cycles. The peaks are rather broad and the $n=3$ peak appears even to be split.
The inset in Fig.~\ref{didval} displays the temperature dependence of the $2\Delta/e$ (i.e. $n=1$) peak. The peak voltages nicely match the BCS curve with a slightly reduced gap.
Being governed by $L_\phi(T)$,\cite{kutch97} the amplitude of the MAR features shows a relatively weak temperature dependence. This is in contrast to the supercurrent which strongly varies with temperature as expected from the exponential dependence of the Josephson coupling on $L_T$. Nearly identical observations have been made on Al/Au/Al junctions. 

In order to further check that the peaks in $dI/dV$ are indeed related to the gap energy we measured another sample with different wire resistance. The critical current of the more resistive sample ($R_N = 34\,\Omega$) was substantially smaller but the peak voltages remained unaffected as demonstrated by the open symbols in the inset in Fig.~\ref{didval}.
The value of the gap \mbox{$\Delta(T=0) =\,163\,\mu$eV} extracted from the $2\Delta$-peaks (see the inset in Fig.~\ref{didval}), is slightly reduced with respect to the bulk value of \mbox{$186\,\mu$eV}.
Earlier experiments on conventional Nb/Nb point contacts\cite{flensb89} have shown a similar suppression of the order parameter at the $n=1$ peak which was attributed to a reduction of $\Delta$ by the relatively high currents which are required to generate the voltage $2\Delta/e$ in low ohmic contacts with 
$R_N \simeq 20-$\mbox{$40\Omega$}. This also leads to deviations from the scaling of the peak voltages, i.e., $V(n=1)/V(n=2)\approx 1.6$ instead of $2$ in Fig.~\ref{didval}. Similar effects are also visible in the data of Ref.~\onlinecite{kutch97}.  

Another important quantity is the excess current $I_{exc}=I(V)-V/R_N$, i.e.~the enhancement of the $IV$-characteristic above the ohmic straight line.  $I_{exc}$ quantifies the integrated proximity correction to $dI/dV$ and saturates at large bias voltages $eV>2\Delta$, where the Andreev reflection is suppressed.
For superconducting point contacts with \mbox{$E_c \gg \Delta$} the excess current is predicted to be $I_{exc}=(\pi^2/4-1)\,\Delta/eR_N\simeq$ \mbox{11$\,\mu$A}.\cite{bardas97} 
In the opposite limit of long diffusive junctions with $E_c\ll\Delta$, $I_{exc}$ is suppressed with increasing length as $1/L$ and amounts\cite{volkov93} to $I_{exc}=0.82\,\Delta/eR_N\cdot\xi^*/L\,\simeq$ \mbox{$2.9\,\mu$A} where $\xi^*=\sqrt{\hbar D_S/\Delta}$ and $D_S\simeq$ \mbox{400 cm$^2$/s}.\cite{note} When integrating the $dI/dV$ curves in Fig.~\ref{didval} we find an asymptotic value of $I_{exc}\,\simeq$ \mbox{$3.5\,\mu$A}, which is in acceptable agreement with the theoretical value obtained in the diffusive limit. The excess current is another feature, which is not contained in the RSJ model.

\begin{figure}[p]
\centerline{\psfig{file=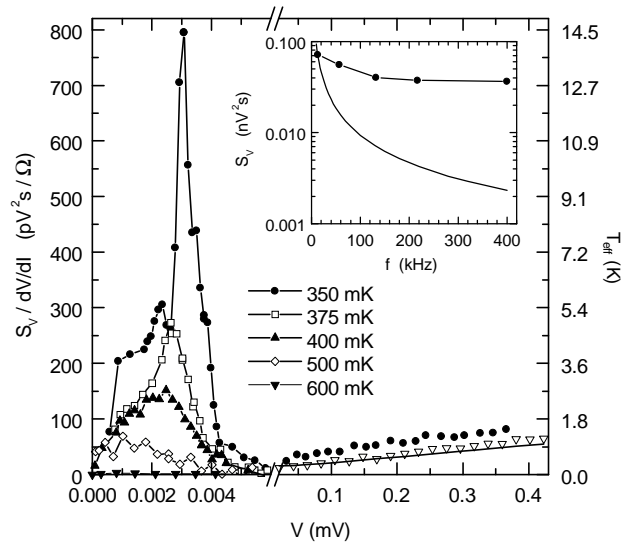,width=8.2cm}}
\caption{\label{sial} Scaled excess noise $S_V / dV/dI$ as a function of voltage for the same device as in Fig.~\ref{didval} with superconducting ($\bullet $) and normal ($\bigtriangledown$) reservoirs. The resistance per junction is \mbox{19.6 $\Omega$} in the normal state. The solid line indicates the shot noise of noninteracting electrons in case of normal reservoirs. 
Note the expanded scale at low $V$, with data taken at $T$ = 350, 375, 400, 500 and 600 mK (from top to bottom). Inset: Amplitude of the low voltage noise peak of a similar sample as a function of frequency. The solid line corresponds to a $1/f$ - dependence.} 
\end{figure}

In the case of coherent MAR it is interesting to check for the existence of multiple charge $q^\ast =\, 2\Delta/V$ transferred during the MAR cycle, which should result in an enhanced shot noise $S_I=2 q^\ast I$ at low voltages.\cite{aver96,cuev98,naveh99}  A first indication for such an effect was seen in NbN based pinhole junctions.\cite{diele97} In Fig.~\ref{sial} we present noise data for the same sample as in Figs.~\ref{rtal} and \ref{didval}. 
We indeed find a huge peak in $S_V$ at very low voltages around 3-4 $\mu$V $\approx 0.02\, \Delta/e$ which vanishes at elevated temperatures together with the supercurrent.

The noise enhancement appears in the strongly nonlinear part of $I(V)$ (inset in Fig.~\ref{rtal}). The measured noise is frequency independent between 100 and \mbox{400 kHz} (see inset in Fig.~\ref{sial}). The nonmonotonic dependence of $S_V$ 

\noindent on $V$ is in strong contrast to the \mbox{2 $\mu$m} long Nb/Au sample of Fig.~\ref{sinb}, which shows no supercurrent at our lowest temperatures and where $S_V$ is monotonically rising with $V$. At higher voltages we find an enhancement of $S_V$ for superconducting reservoirs (full circles) with respect to normal reservoirs (open triangles). In the voltage regime $V \gtrsim\Delta$ the noise enhancement is most likely caused by heating similar to the Nb/Au case discussed above, where the heating is even more pronounced because of the larger gap of the Nb.   

\section{Discussion}
One possible origin of the low voltage noise peak are temporal fluctuations of the critical current as previously observed in grain boundary junctions made from high temperature superconductors.\cite{marx95} Such critical current fluctuations may be caused by the motion of localized defects close to the junction and should result in a  $1/f$ - like frequency dependence of the voltage noise close to $I_c$ as well as of the normal state resistance $R_N$. The latter would result in a parabolic increase of $S_V$ for $I>I_c$ which is absent in Fig.~\ref{sial}. 
At our typical measuring frequencies \mbox{$f>100\,$kHz} the measured peak height is independent of $f$ (see the inset in Fig.~\ref{sial}). For \mbox{$f < 100\,$kHz} we observe a small increase of the peak amplitude which is currently not understood, but certainly inconsistent with a $1/f$ law.
Hence, $1/f$ noise can be ruled out as the origin of the low voltage noise peak.

Earlier experiments on shunted tunnel junctions have also revealed an increase of the noise at low voltages.\cite{koch82} This effect has been predicted\cite{seme72} to arise as a consequence of Johnson-Nyquist noise of the shunt resistor. Fluctuations at high frequencies are mixed down to low frequencies by the highly nonlinear $IV$ characteristics of the junction. Good agreement with the experiment has been found for both the noise rounding of the $IV$ curves and the excess noise. 
When calculating the noise according to the RSJ model using the measured $dI/dV$ and $I_c(T)$ in the low voltage region for the temperatures shown in Fig.~\ref{sial} we find a peak with an amplitude of \mbox{35 pV$^2$s$/\Omega$} at \mbox{350 mK} which is about 20 times smaller than the measured noise peak.

We believe that the noise peak is related to a strongly fluctuating supercurrent at the onset of finite voltage. As discussed already in the context of the $IV$ curves in Fig.~\ref{rtal}, temporal fluctuations of $|F(x)|$ can be thermally excited at the weak spot in the center of the N wire. These lead to large fluctuations of the supercurrent, and consequently to both large noise and unusually broad $IV$ curves. 
The minimum in $|F(x)|$ is the specific feature of junctions longer than $L_T$ and is not contained in the treatment of Refs.~\onlinecite{aver96,cuev98,naveh99}. 

The thermally activated fluctuations of the supercurrent have to be distinguished from the fluctuations of the critical current discussed above. The latter correspond to fluctuations of the activation energy with an $1/f$ spectrum, which are negligible at the time scale of $\mu$s, where our noise measurements are usually performed.

Independent support of this interpretation is provided by the recent observation of Thomas {\it et al.},\cite{thomas98} who found a similar thermally activated rounding of the $IV$ characteristics in InAs-based SNS junctions. Their samples are also in the regime $L>L_T$ and the measured activation energy is typically two orders of magnitude smaller than expected from the RSJ model.  
In our samples, $R(T)$ is also broader than expected from the RSJ model (see the inset in Fig.~\ref{rtal}).
Thomas {\it et al.} suggest that the rounding of the $IV$ characteristics may be caused by an additional current noise, which is much larger than the Johnson noise of the device. Our work provides direct experimental evidence for such an enhanced noise at the onset of Josephson coupling in "long" ($L>L_T$) SNS contacts.

\begin{figure}[p]
\centerline{\psfig{file=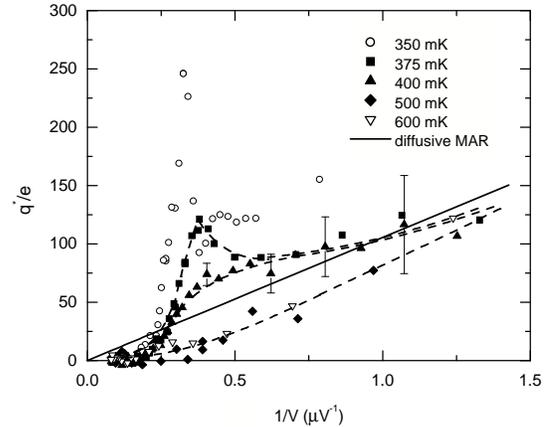,width=7.4cm}}
\caption[]{\label{qstar}  Effective multiple charge $q^* = S_I/2eI$ as a function of $1/V$ corresponding to the data in Fig.~\ref{sial}. The solid line indicates the theoretical estimate for  $q^*$ for the diffusive case.\cite{naveh99} The dashed lines are a guide to the eye. The error bars indicate the uncertainty due to the subtraction of the background noise.}
\end{figure}

In contrast to the supercurrent the MAR induced subharmonic gap structure in $dI/dV$ is much less temperature dependent (see Fig.~\ref{didval}).
The most striking signature of higher order MARs would be the presence of shot noise of multiple charge quanta. 
In order to check for MAR induced low voltage noise we have to look at higher temperatures where the supercurrent and its corresponding noise peak are suppressed.
In Fig.~\ref{qstar} we plot the effective charge $q^*/e=S_I/2eI$ vs.~$1/V$.  At low temperatures $T\lesssim$ \mbox{500 mK} the noise peak in $S_V$ is reflected also in $q^*$. Remarkably, the noise raises again for even lower voltages, instead of dropping to zero as expected if the supercurrent noise is the only noise source.
At higher temperatures $T\gtrsim$ \mbox{500 mK} the peak associated to supercurrent fluctuations vanishes, but we still find a noise signal which rises roughly linear with $1/V$. At the lowest voltages all curves (dashed lines) seem to merge into a straight line with a slope only slightly lower than $\approx 0.3 \cdot 2 \Delta$ as predicted by the theory for the diffusive regime (solid line).\cite{naveh99} 
The theory by Naveh and Averin for the MAR noise considers very short junctions with $E_c \gg \Delta$. In our long SNS junctions $E_c \ll \Delta$ is the {\it smallest} energy scale. To our knowledge, the shot noise has not yet been calculated for this case. 

The measured effective charge ranges up to 100 $e$, which is suprisingly large since the coherence of the MAR cycle is expected to be cut off by inelastic scattering in our samples after a few Andreev reflections. From this point of view, it is already surprising that we find up to four MAR peaks in $dI/dV$. This raises the question whether phase coherence over $n \times L$ ($n$ is the number of Andreev reflections) is required  or only over 1 or 2 $\times$ $L$.  Although the magnitude and the functional dependence of the low voltage noise in Fig.~\ref{qstar} are compatible with the existence of multiple charges, we cannot exclude other possibilities. Further experiments are required to separate the contributions from the supercurrent noise and the possible shot noise of multiple charge quanta.

\section{Conclusions}
By means of noise measurements we have shown that multiple Andreev reflections in a normal metal wire sandwiched between two superconductors lead to substantial  electron heating in the wire. The strength of this heating effect depends on the size of the gap in the superconductors. For Nb with a large gap the effective electron temperature 
raises already for small currents up to several K, which leads to a suppression of coherent multiple Andreev reflection.  For the smaller gap superconductor Al the heating is less pronounced and phase sensitive effects such as subharmonic gap structure become visible. With the onset of a proximity induced supercurrent through the normal wire a sharp noise peak appears at low voltages, which we attribute to thermally induced fluctuations of the supercurrent. When the supercurrent is suppressed at moderately elevated temperatures an additive noise contribution remains at low voltages, which suggests the existence of multiple charge quanta with charge much larger than $e$.

\section{Acknowledgements}
We acknowledge helpful discussions with N.~Argaman, D.~Averin, C.~ Bruder, H.~ Kroemer, Y.~Naveh,  H.~Pothier, B.~Spivak, E. Sukhorukov, and B.~van Wees. This work was supported by the Swiss National Science Foundation.
\vspace{-.0cm}


\begin{thebibliography}{10}
\bibitem{note}
The coherence length on the superconducting side enters because of the modulation of the order parameter of the superconductor at the NS interface.
\end{thebibliography}


\end{document}